\documentclass[conference]{IEEEtran}
\IEEEoverridecommandlockouts

\usepackage{multirow}
\usepackage{cite}
\usepackage{amsmath,amssymb,amsfonts}
\usepackage{algorithmic}
\usepackage{graphicx}
\usepackage{textcomp}
\usepackage{xcolor}
\usepackage{balance}
\def\BibTeX{{\rm B\kern-.05em{\sc i\kern-.025em b}\kern-.08em
    T\kern-.1667em\lower.7ex\hbox{E}\kern-.125emX}}
\begin{document}

\title{Impacts of the Space Technology Evolution in the V\&V of Embedded Software-Intensive Systems}

\author{\IEEEauthorblockN {Carlos L. G. Batista\IEEEauthorrefmark{1},
Tania Basso\IEEEauthorrefmark{2},
F\'atima Mattiello-Francisco\IEEEauthorrefmark{1},
Regina Moraes\IEEEauthorrefmark{2} \IEEEauthorrefmark{3}
\IEEEauthorblockA{\IEEEauthorrefmark{1}
National Institute for Space Research (INPE), S\~ao Jos\'e dos Campos, Brazil\\ carlos.batista, fatima.mattiello \{@inpe.br\}}
\IEEEauthorblockA{\IEEEauthorrefmark{2}
University of Campinas (UNICAMP),
Limeira, Brazil\\taniabasso, regina 
\{@ft.unicamp.br\}} \IEEEauthorblockA{\IEEEauthorrefmark{3} University of Coimbra (UC), Coimbra, Portugal - Full / Regular Research Paper  (CSCI - ISSE)} }
}
\maketitle
\begin{abstract}

CubeSat-based nanosatellites are composed of COTS components and rely on its structure and standardized interfaces. A challenge in the nanosatellites context is to adapt the V\&V (Verification and Validation) process to answer to the increase importance of the embedded software, to reduce the artefacts to be delivered aiming at cutting cost and time and still complying with international standards. This work presents an analysis of the strategy adopted in a real nanosatellite for the development of the OBDH software embedded in NanosatC-BR2 mission. The goal is to discuss the impact that the standardization of the structure and interfaces of the CubeSat impose on the V\&V process of the SiS and to highlight the challenges of “New Space Age” for the use of existing V\&V techniques and methods.

\end{abstract}


\section{Introduction}

In the last decade the space area has evolved significantly, thanks for the advent of microprocessors and the spread of software as a service that allowed the development of micro and nanosatellites programs. The advent of the CubeSat standard has made the development cycle of small satellite projects (nanosatellites) much faster and cheaper, compared to larger space missions. In addition to the use of COTS (components off-the-shelf), that reduce cost, the standardization of the satellite structure and its interfaces helps improving the architectural satellite solutions based on CubeSat \cite{twiggs2008origin}. 

With the technological evolution of embedded electronics, memories and satellites processors, the functionalities implemented by software have increased.
The so-called software-intensive systems (SiS) embedded in satellites are increasing in terms of complexity and becoming critical elements for the mission.
Moreover, the “new space age”, i.e., greater access to space by different segments of the society, demands for new space products and services, requiring regulation and well-established development processes. CubeSat can be taken by an example due to its popularity within the academia and industry. The complexity of CubeSat structure, cabling and interfaces has been significantly simplified with the standardization, but the software has not. 

Addressing this problem, SiS suppliers are reducing both the design artifacts to be delivered in the reviews baseline and the number of reviews in the development cycle of many SiS embedded in nanosatellites. However, negative side effects of this strategy in the quality of the nanosatellite V\&V process (unknown software requirements in the begin of the development cycle, immaturity of the stakeholders and expertise of the development time in the space area projects) have been observed regarding the lack of compliance with international standards.
The challenge is to establish effective V\&V approaches that can contribute to make the nanosatellites V\&V process shorter and cheaper but mainly compliant with the international standards. 


This work presents the V\&V strategy adopted in the development of NanosatC-Br2,  a scientific nanosatellite leading by  the Brazilian Institute for Space Research (INPE) with partners in Brazilian Universities and software private supplier. Considered the main SiS element of the mission, and the focus of this study, the On-Board Data Handling (OBDH) software is responsible to the implementation of the interfaces with the mission payload and the ground systems. The approach addresses the V\&V activities with the respective techniques and methodologies adopted by INPE in the development of  SiS projects embedded in satellites based on the guidelines of the European Cooperation for Space Standardization (ECSS) \cite{ECSS-E-ST-10C}
The V\&V activities rely on baseline reviews and Model-Based Testing (MBT) to generate, automatically, test cases, being able to systematize the OBDH testing. 



In addition to the introduction, Section \ref{sec:brackground} presents the background and related work. Section \ref{sec:cubesat} presents the  V\&V process adopted by INPE in the development of OBDH SiS. Section \ref{sec:results} presents research opportunities to improve the  V\&V process of standard Cubesat satellites, ending with the conclusions and future work in Section \ref{sec:conclusions}.

\section{Background and related work}
\label{sec:brackground}

The CubeSat standard emerged from an effort by California Polytechnic State University and Stanford University in 1999. It is a simple and small satellite model, developed with low-cost (less than one million Euro) and a short project cycle (less than two years to flight readiness) educational objectives, both in development and in launch \cite {twiggs2008origin}. It is a cube-shaped nanosatellite, whose edges measure 10 centimeters, volume of one liter and can hold a bus and payload with mass of 1.3 kilograms. More than one unit (or 1U) can be combined to form larger satellites (e.g. 2U, 8U) and can use ejection systems in standardized orbits (e.g. P-POD - Poly Picosatellite Orbital Deployer), allowing to release several satellites through the same interface and to launch several CubeSat in a single launch vehicle or share trips with supplies in the refueling operations from ISS (International Space Station). 

\begin{figure}[ht]
	\centerline{\includegraphics [width=.80\linewidth]{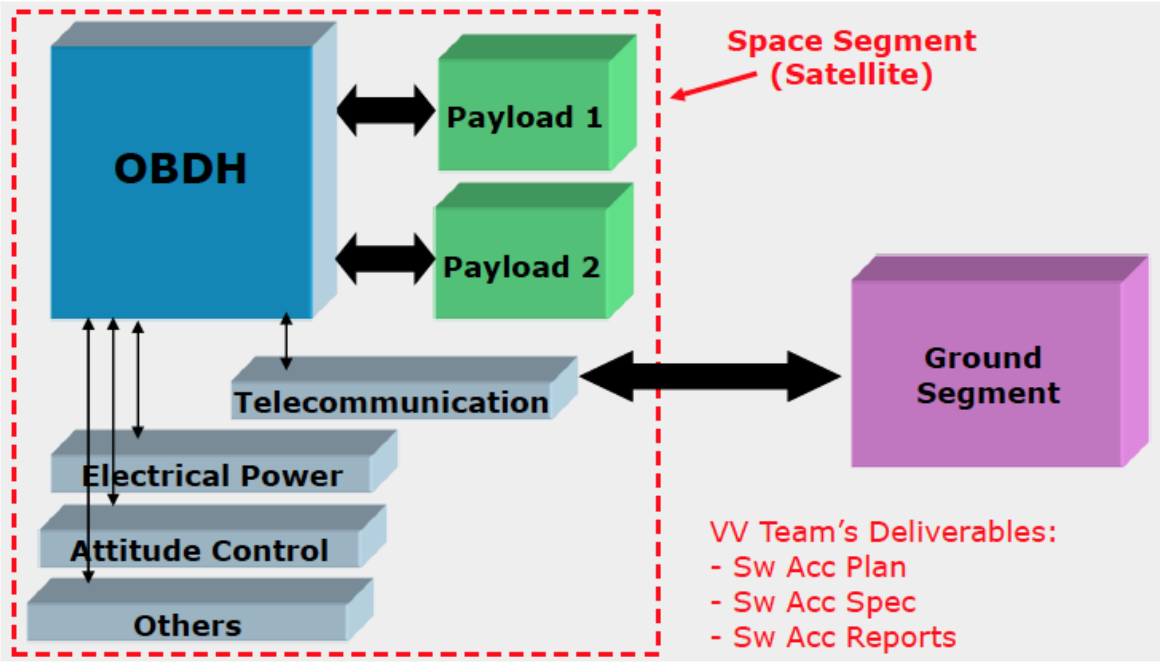}}
	\caption{Architecture of the space and ground segments}
	\label{fig:architecture}
\end{figure}

The architecture of a satellite mission is composed of four parts: the space, the launch, the ground and the user segments. The space segment is the more important one for this work. Within it are the payload (set of equipment dedicated to the application of the space mission) and the platform (set of subsystems designed to support the operation of the mission in orbit, also referred as \textit{bus}). Figure \ref{fig:architecture} shows the typical architecture of a satellite platform, with the space segment subsystems and its communication with the ground segment. We highlight the OBDH (On-Board Data Handling) subsystem, a typical SiS, which is responsible for monitoring the health of the other subsystems of the satellite platform, for interacting with the payloads of the mission and for ensuring communication with the ground in the operation of the mission.

Besides dataprocessing, OBDH plays the role of the spacecraft \textit{command \& control} and \textit{telemetry data management}. 

The \textit{command \& control} functions on satellites enable remote operation of the satellite subsystems, allowing the ground controller to change the behavior of the spacecraft as, for example, change the operational mode and coordinate orbital maneuvers.

The functions associated with \textit{telemetry} on board include real-time monitoring of the platform's equipment and subsystems and also of the payload in terms of critical parameters to the mission's survival. The data packaging and temporary storage of mission on board are also functions performed by the OBDH software, responsible to transmit those data to the ground in the window of visibility of the satellite and Ground Stations. The platform telemetry generally encompasses two categories of data: \textit{housekeeping} (parameters needed to check the health and state of the spacecraft) and attitude parameters. The payload telemetry consists of the application's data set (the mission's data). Computing on board of satellites is also used to reduce the telemetry, such as compressing data on board to reduce the amount of data and reduce the needs of high downlink bit rates.


Software engineering is normally addressed in the space project standards. Best practices are recommended by NASA and ESA in cooperation with the industry \cite{ECSS-E-ST-10C}\cite{nasa2016handbook} . 
In fact, standardization is important to reduce risks, costs and improve quality and communication in the space for any spacecraft mission, including CubeSat \cite{twiggs2008origin}. Scholz \cite{scholz2017cubesat} presents a standard handbook reviewing existing space standards and their potential application within CubeSat mission.


Focused on the V\&V process, the interests of this work are on techniques and tools to assure the quality of the software embedded in the spacecraft. The most frequent techniques being applied to improve and assure quality in space software are reviews and tests. According to ECSS standard \cite{ECSS-E-40}, 
reviews assure good quality of documents and are a good opportunity for customers to check, gradually, the comprehension of the suppliers about the software details. The tests demonstrate that the implementation meets the requirements and performance constraints are correctly and completely implemented.

\begin{figure}[ht]
	\centerline{\includegraphics [width=.99\linewidth]{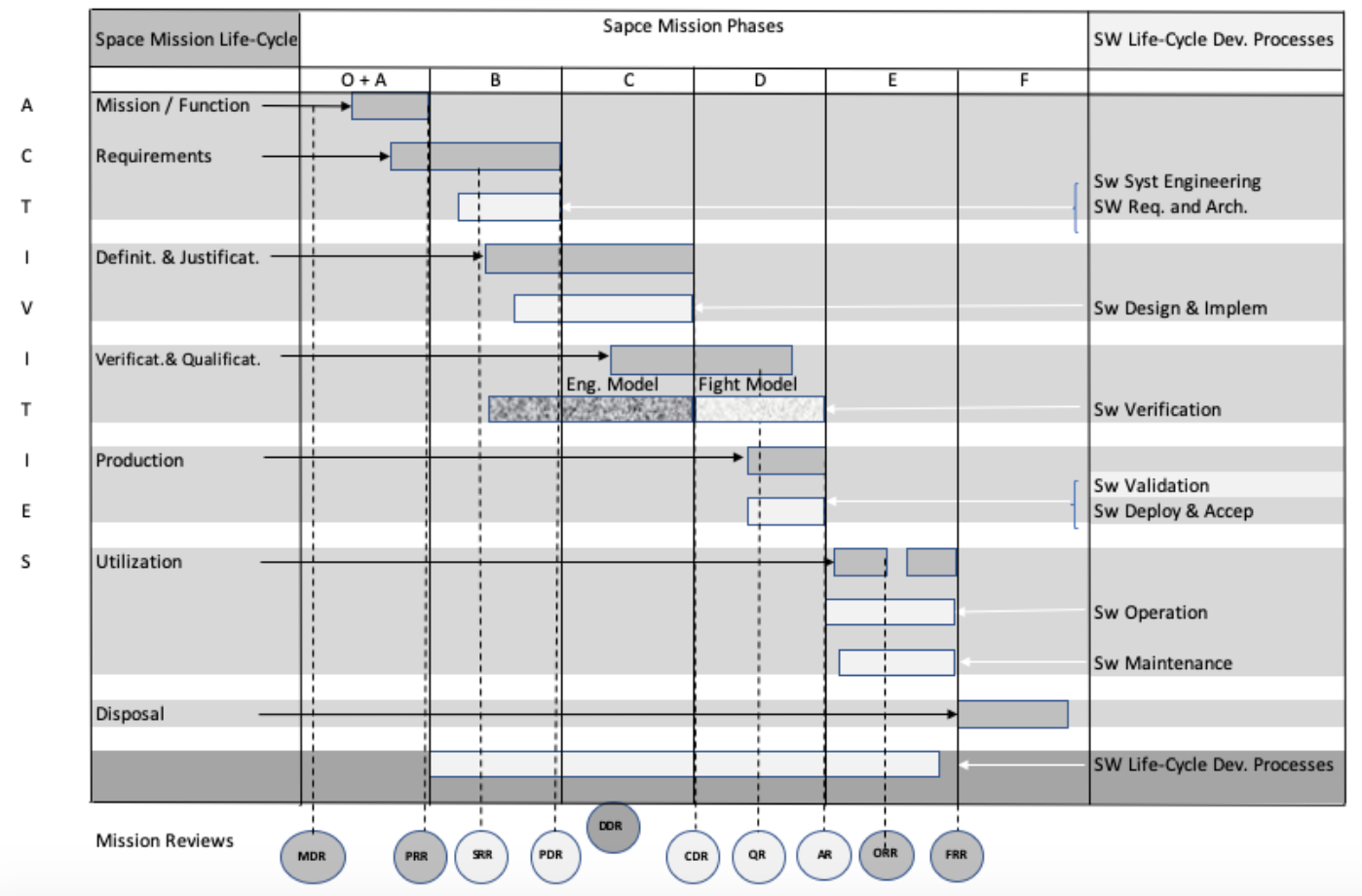}}
	\caption{ECSS Space Mission Project Life Cycle}
	\label{fig:mission}
\end{figure}

Figure \ref{fig:mission} presents the basic mission (column 1) and related software (rightest column) life cycle activities recommended in ECSS, along the typical space mission phases: Mission Analysis/ Needs Identification (O), Feasibility (A), Preliminary Definition (Project and Product)(B), Detailed Definition (Product) (C), Production/Ground Qualification Testing (D), Utilization (E), Disposal (F) depicted in the central part of this figure 
. The darker bars represent the period that each space mission activity is carried out under system point of view. The lighter bars correspond to the software activities related to the software life-cycle development under software subsystem point of view (i.e., phases B, C and D). 

On the bottom of Figure \ref{fig:mission}, can be observed the sequence of reviews in the space mission project, at system engineering level. They are: MDR - Mission Definition Review, PRR - Preliminary Requirements Review, SRR - System Requirement Review, PDR - Preliminary Design Review, DDR - Detailed Design Review, CDR – Critical Design Review, QR - Qualification Review, AR - Acceptance Review, ORR - Operation Requirements Review, FRR - Flight Readiness Review. The software project (a subsystem in the hierarchy of space system development) shall be synchronized with those milestones.

Along the life cycle a series of models must be elected to help the V\&V processes. Three main physical models of the satellite are considered: (i) Engineering Model (EM) - the development environment based on prototyping with non-qualified components; (ii) Qualification Model (QM) - the satellite integrated in the flight structure to be stressed on testing; and (iii) Flight Model (FM) - the flight structure assembled for flying. From a software project perspective QM and FM are considered identical.  Note the bar associated with the software verification activities. It considers the software at CDR is already embedded in the hardware, being the OBDH subsystem (software and hardware integrated).

In CubeSats, the OBDH grows in complexity as long as the hardware gets miniaturized. More and more services are delegated to the on board software replacing functions once done by hardware. More pieces of software are developed to interface, control and operate different subsystems. These new features increase the overall complexity of the satellite and also increase the number of vulnerabilities 
. The success of the mission relies on the perfect operation of the OBDH 
, even more for CubeSats.

Different works have been proposing solutions to mitigate vulnerabilities without increase the costs of a V\&V process. Some focusing on specific pieces of software \cite{kiesbye2019hardware}\cite{leppinen2019developing}, others new methodologies \cite{alanazi2019engineering}\cite{venturini2018improving}, others to develop models in model-driven approaches \cite{eickhoff2003mdvv}\cite{khan2012model}, and, even more, some of them exploit the standards tailoring \cite{tiseo2019tailoring}\cite{feldt2010vv}.

\section{V\&V process for OBDH developed at INPE}
\label{sec:cubesat}
CubeSat establishes in its minimum structure (1U) electrical interfaces with standardized connectors through which the subsystems of the satellite platform connect themselves (\textit{bus}). In this way, the architectural complexity of the satellite in terms of structure and cabling becomes significantly smaller, adding the advantage of several COTS subsystems that the industry is able to supply.

Due to increased demand of nanosatellite mission for the purpose of new technology demonstration in orbit, in 2016, the European Space Agency (ESA) released a document entitled \textit{Tailored ECSS Engineering Standard for In-Orbit Demonstration (IOD) CubeSat Projects}\footnote{https://copernicus-masters.com/wp-content/uploads/2017/03/IOD\_CubeSat\_\\ECSS\_Eng\_Tailoring\_Iss1\_Rev3.pdf}. IOD Cubesats are considered structure (1-U or multiple) generally characterized by the following attributes:
(i) Complete stand-alone systems including platform, payload, ground segment and operations; (ii) Higher risk acceptance profile; (iii) Low level of complexity (relative to other space ESA project); (iv) Low cost and short schedule; (v) Short operational life-cycle (less than one year); (vi) Acceptance of single point of failures; (vii) Limited redundancy and fault tolerance; (viii) Robust safe mode; (ix) Extensive use of COTS elements and extensive tests focusing on system level (functional and environment, qualification and acceptance); (x) Simple project organization with well integrated teams (few suppliers or subcontractors).

ESA states that 
only a few in the list of classical acquisition standards may be selected as applicable. Thus, ``The IOD verification process shall be adapted to reflect the reduced complexity of work and the absence of one or more disciplines". Regarding software discipline, INPE has invested in reducing complexity of the verification process of OBDH software developed for being used in two ongoing Cubesat-based missions, named NanosatC-Br2 and CONASAT-0.

\begin{figure}[ht]
	\centerline{\includegraphics [width=.85\linewidth]{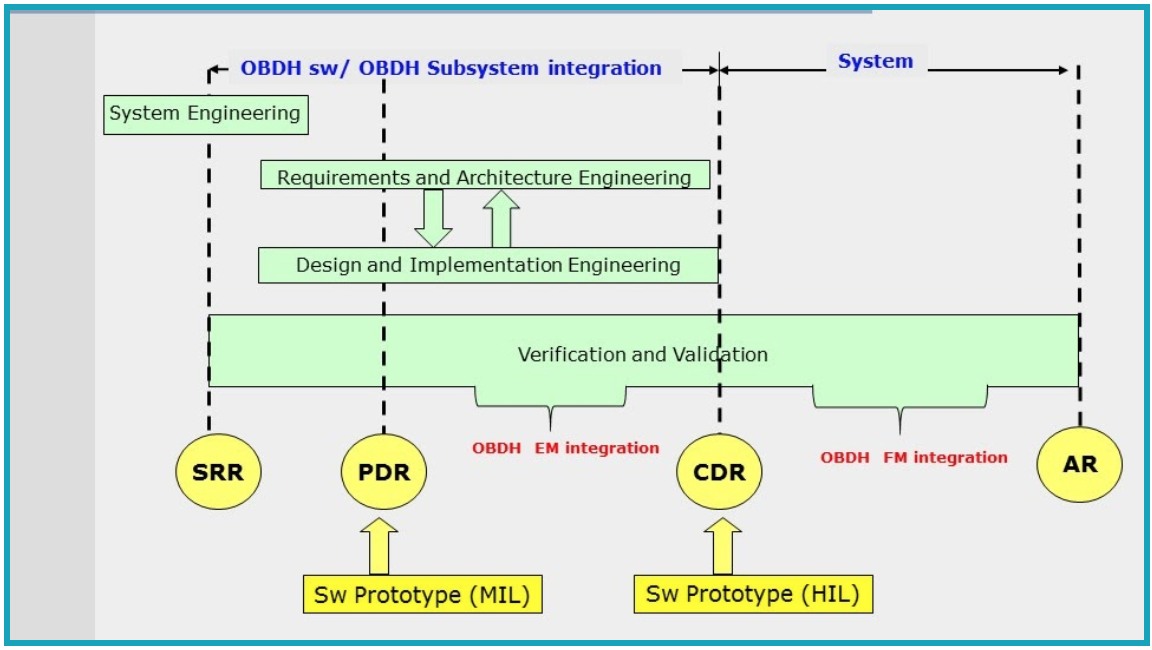}}
	\caption{Software Development Life Cycle for Nanosatellites}
	\label{fig:swlifecycle}
\end{figure}

The V\&V process is still supported by reviews at design time as recommended in ECSS. However, in the light of the ECSS Mission Project Life Cycle, already presented in Figure \ref{fig:mission} and fully implemented for the traditional satellites, CubeSat project has eliminated the revisions \textit{DDR - Detailed Design Review} and \textit{QR - Qualification Review}. The justification is that DDR aims, in traditional satellites, to verify with the team hired for the development, the consistency of the design of the drivers and software libraries with the hardware also under development. In the case of Cubesat nanosatellite, the use of standardized interfaces and protocols available as third-party components (COTS) guarantees the necessary compatibility. Similarly, since the basic Cubesat 1-U platform is already provided as a qualified COTS, QR review that is justified on traditional satellites to demonstrate that the satellite structure (QM) with its integrated subsystems has been verified and can be manufactured for flight (FM) is no more necessary. However, the AR - Acceptance Review was maintained, as it is a system-level review.
Figure \ref{fig:swlifecycle} shows the adaptation of the ECSS Mission Software Life Cycle carried out at INPE for nanosatellites. The standardization of hardware benefits the embedded software design due to the availability, during its development, of COTS on-board computer architectures where the software should be embedded. 

Thus, with the help of Model Driven Engineering (MDE) approaches, early in the development cycle, it is possible to support the requirements definition using techniques that allow to verify the behavior of the embedded software as expected by means executable models (Model-in-the-loop - MIL) in the PDR revision and effectively embedded in the target hardware (Hardware-in-the-loop - HIL) in the CDR revision (see Figure \ref{fig:swlifecycle}). MIL prototyping makes it possible to anticipate, in the mission development cycle, the identification of SiS interoperability and robustness problems that might occur on the interaction of two communicatiing SiS \cite{approach} 
.

\subsection{Deliverables for Mission Reviews}
\label{sec:deliverablesCube}
This section presents an example of a typical V\&V process tailored by INPE for the development of an On-Board Data Handling (OBDH) subsystem embedded into Cubesat-based satellite missions. The process outputs include both a set of deliverables considered input for each mission review in Figure \ref{fig:swlifecycle} and the testing method used for the integration of the OBDH with other subsystems of the platform, and other SiS such as the satellite payloads.

The deliverables related to each review of the software development life cycle for Cubesat-based missions are in the third column of Figure \ref{tab:reviewDeliverables}. It is shown in bold the final delivery version and the respective reviews are the ones in which the document is considered closed.  \textit{Requirement Baseline} and \textit{Interface Requirements} are not closed at the end of SRR review, as in traditional satellites, but only at the end of CDR. This is feasible in software designing that follows prototyping and incremental approaches.

\begin{table}[ht]
\caption{OBDH (SiS) Mission Review and Deliverables}
\label{tab:reviewDeliverables}
\resizebox{\linewidth}{!}{%
\begin{tabular}{c|l|l|l}
\hline
\multicolumn{1}{l|}{\textbf{Mission Reviews}} & \textbf{Deriverables - Traditional Satellite} & \textbf{Deriverables - CubeSats} & \textbf{Deriverable Owner} \\ \hline
\multirow{4}{*}{SRR}                          & \textbf{Requirements Baseline}                & Requirements Baseline            & Customer                   \\
                      & \textbf{Interface Requirements}    & Interface Requirements             & Customer  \\
                      & SW Development Plan                &                                    & Supplier  \\
                      & V\&V Plan                          & V\&V Plan                          & V\&V Team \\ \hline
\multirow{6}{*}{PDR}  & \textbf{SW Development Plan}       & SW Design                          & Supplier  \\
                      & \textbf{Technical Specification}   & Requirements Baseline              & Supplier  \\
                      & V\&V Plan                          &                                    & V\&V Team \\
                      &                                    & Interface Requirements             & Customer  \\
                      &                                    & SW Test Plan                       & Supplier  \\
                      &                                    & SW Test Report                     & Supplier  \\ \hline
\multirow{12}{*}{CDR} & \textbf{SW Design}                 & \textbf{SW Design}                 & Supplier  \\
                      & \textbf{SW Source Code}            & \textbf{SW Source Code}            & Supplier  \\
                      & \textbf{SW Test Plan}              & \textbf{SW Test Plan}              & Supplier  \\
                      & \textbf{SW Test Reports}           & \textbf{SW Test Reports}           & Supplier  \\
                      & User Manual                        & User Manual                        & Supplier  \\
                      & V\&V Plan                          & V\&V Plan                          & V\&V Team \\
                      & V\&V Plan Instrument               & System AIT Plan                    & V\&V Team \\
                      & V\&V Specification                 & Mission Operation Concept          & V\&V Team \\
                      & V\&V Plan Subsystem                &                                    & V\&V Team \\
                      & V\&V Specification Subsystem       &                                    & V\&V Team \\
                      &                                    & \textbf{Requirements Baseline}     & Customer  \\
                      &                                    & \textbf{Interface Requirements}    & Customer  \\ \hline
\multirow{5}{*}{AR}   & \textbf{SW approval}               & \textbf{SW approval}               & Customer  \\
                      & \textbf{User Manual}               & \textbf{User Manual}               & Supplier  \\
                      & \textbf{V\&V Plan System}          & \textbf{System AIT Plan}           & V\&V Team \\
                      & \textbf{V\&V Specification System} & \textbf{System AIT Report}         & V\&V Team \\
                      & \textbf{V\&V Report System}        & \textbf{Mission Operation Concept} & V\&V Team \\ \hline
\end{tabular}%
}
\end{table}


\subsection{MIL and HIL prototyping for SiS integration purpose }

\begin{figure}[ht]
	\centerline{\includegraphics [width=.85\linewidth]{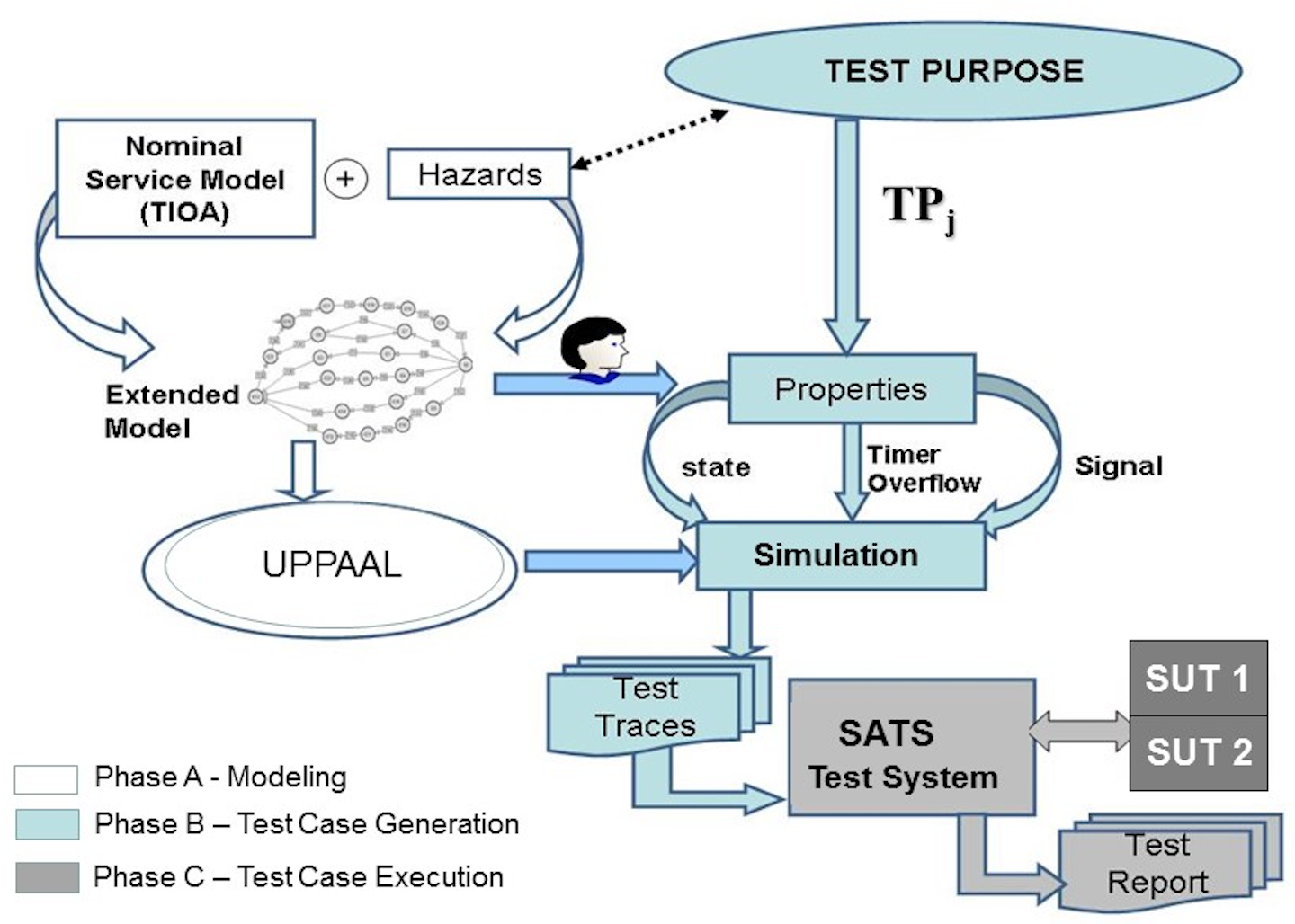}}
	\caption{InRob Testing Process}
	\label{fig:inrob}
\end{figure}

Inspired by the antecipation of issues related to OBDH SiS integration with the nanosatC-BR2 payloads, the InRob testing process originally proposed in  \cite{mattiello2012inrob} and its extension in InRob-UML \cite{weller2015inrob} have been used to guide the construction of communicating models (MIL). These models represent the expected behaviour when two communicating SiS perform a given service collaboratively. The models execution allows to both verify  interoperability and robustness requirements by simulating and deriving tests cases using MBT tools.
Figure \ref{fig:inrob} presents InRob Testing Process in three phases: A) Modeling; B) Test case Generation; C) Test case Execution.

The inputs for the service modeling (phase A) are: (i) the project artifacts that specify among others the system functional requirements, the subsystem user manual and time constraints; (ii) the service profile development, which supports the prioritization of a particular service to be modeled. 
The nominal behavior of the elected service is formally specified. An interoperability TIOA model results in the nominal TIOA model. From the nominal service model and time constraints imposed on the communicating SIS, the concept of Major and / or Minor timing deviations are used to deal with  timing hazards related to the communication channel failures. Thus, the nominal service model is extended to represent the robust behavior of each communicating SIS, facing those considered deviations. The output is the extended model of the service. 

The extended TIOA models outputs in Phase A are the basis for the test cases generation methods oriented to test purpose (TP) approaches in Phase B. A TP is a formal description of a property (or a sequence of properties) represented in the TIOA models. On-the-fly methods for test case generation search for a specific property during the simulation and the test case is generated by recording a trace on traversing the model aiming at the property verification. This test case is able to validate the behavior of the implementation concerning that property. Robustness test case generation considers hazard scenarios (e.g. rushed and duplicate messages) specified by the test purpose approach. 

In Phase C of the testing process, the test cases were performed in real SUT. The testing architecture relies on a Failure Emulator Mechanism (FEM), which is considered as part of the Testing System (TS).

\begin{figure*}[ht]
	\centerline{\includegraphics [width=1\linewidth]{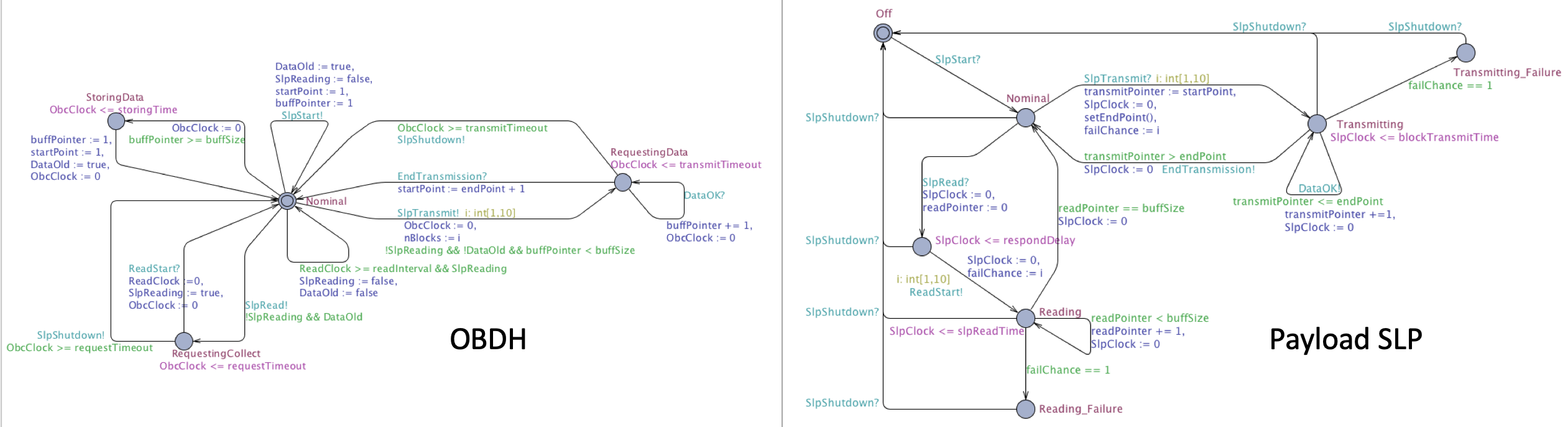}}
	\caption{OBDH and Payload SLP interoperability models}
    \label{fig:obdhPayload}
\end{figure*}

The InRob approach has been used in the V\&V activities of the nanosatellite named NanosatC-Br2, a  scientific 2U Cubesat-based nanosatellite under development at INPE.
NanosatC-BR2 carries on Earth orbit 5 payloads being 2 scientific instruments and 3 new technologies for validation in the space. InRob aids to create representative models for interoperability and robustness analysis of the OBDH expected behavior (requirements) in the  interactions with the nanosatellite payloads. Following InRob Fase A, the modeling was built with the aid of UPPAAL tool, using timed automata formalism, enabling the analysis of the requirements through model simulation.
Ten models in total output from Fase A, being two for each payload. Figure \ref{fig:obdhPayload} shows the nominal models of the interactions between OBDH and payload SLP (Scientific Langmuir Proble payload) that occur on the channel I2C in a master-slave communication. From those models, test cases can be automatically derived through the verification of properties in the simulation of the models using model-checking technique, in Fase B.

To start the reading, the on-board computer will send a command to the reading initialization SLP, which will contain bytes referring to the local date and time (a timestamp to deal with the lack of a Real Time Clock). So, the SLP will print at the beginning of each reading the date and time of the beginning of the collection. The SLP must respond to the OBDH the correct receipt of the command (acknowledge). After starting the reading, the OBDH should estimate a period of more than 5 minutes to start asking for data transmission (avoiding to interrupt the SLP in the collection). If an error occurs and the OBDH requests data transmission to the SLP, the SLP must ignore the request. 

\begin{figure}[ht]
	\centerline{\includegraphics [width=1.00\linewidth]{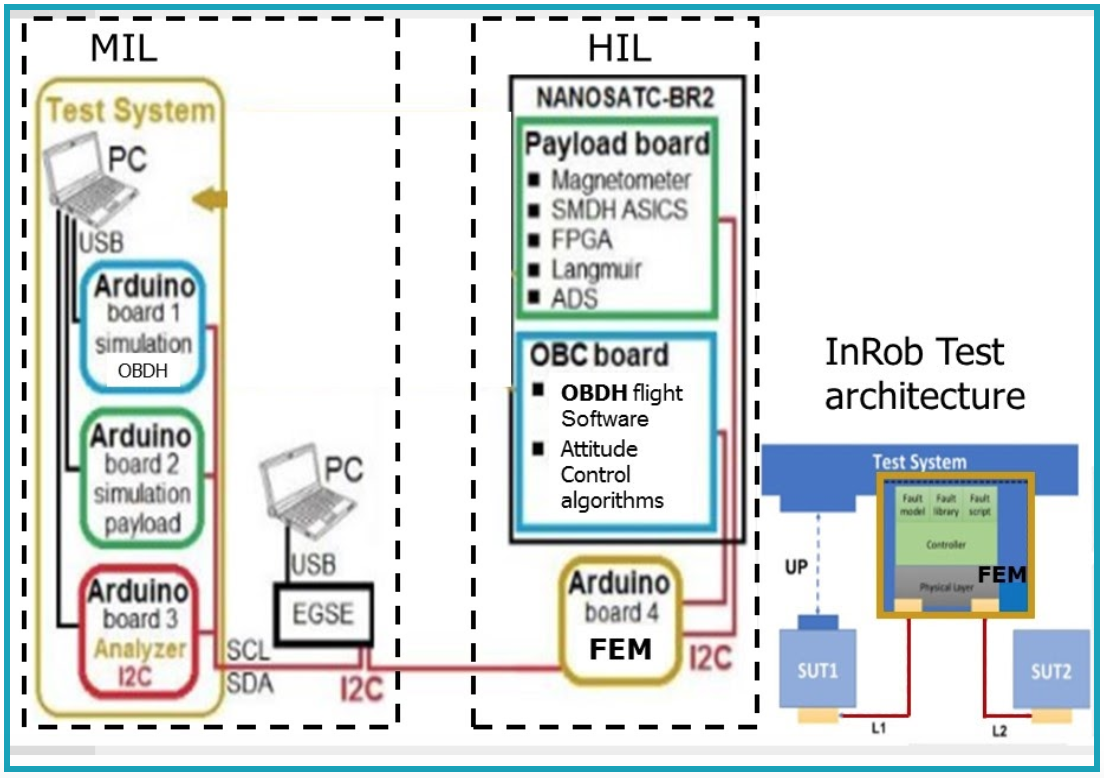}}
	\caption{InRob Test Architecture instantiated by SATS \cite{approach} 
	}
    \label{fig:inrobTest}
\end{figure}

The implementation of Fase C in the NanosatC-BR2 count on a Scalable Architecture Test System (SATS) based on Arduino computer boards proposed in Figure \ref{fig:inrobTest} \cite{approach} 
. SATS supports the V\&V process on the MIL and HIL perspectives. In MIL the codes generated from models presented in Figure \ref{fig:obdhPayload} are embedded in the Arduinos and the interactions between OBDH and SLP can be verified from a behavioral point of view. That facility allows us to anticipate requirement's ambiguities and specification faults. Moreover, functional test cases can be derived from those models. Those test cases will be executed in the HIL environment when the communicating SiS are embedded in the real hardware, for acceptance purposes.

The Failure Emulator Mechanism (FEM) in the InRob Test Architecture (see Figure \ref{fig:inrobTest}) supports faut injections in the communication channel that is used for the SiSs (SUT1 and SUT2) interactions in order to validate robustness requirements. FEM is able to intercept the exchanged messages between two subsystems under testing and inject different faults: (i) time related faults, i.e. delay; (ii) value related faults, i.e. bit-flip and (iii) specific communication bus faults, i.e. verbose subsystem.

\begin{figure}[ht]
	\centerline{\includegraphics [width=1.00\linewidth]{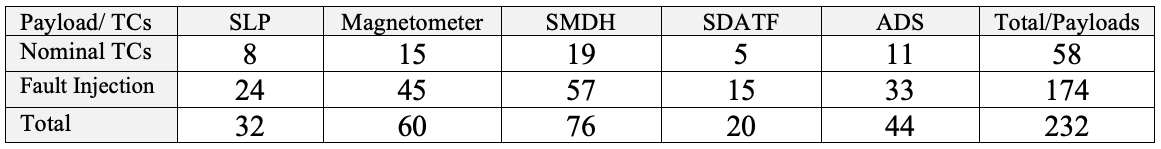}}
	\caption{Number of Test Cases generated by InRob}
    \label{fig:inrobResult}
\end{figure}

Figure \ref{fig:inrobResult} presents the number of Test Cases derived from the ten models generated  with the aid of UPPAAL tool in Fase A. 
For each interaction between the SUTs models of interoperability, which is considered a Test Purpose (TP), one Test Case classified as nominal was automatically generated. Regarding the interoperability models OBDH and SLP, there are 8 interactions specified, therefore 8 nominal TCs. Moreover, for each nominal TC, at least 3 TCs are supplemented by FEM during test execution, associated with the three fault models implemented. In total  32 TCs were executed. 
For the five payloads of NanosatC-Br2, from 10 interoperability models specified using InRob, 232 test cases were automatically created, 58 nominal functional and 174 fault injection test cases (applied by FEM).

The conception of a HIL environment starts in the EM integration as a continuous integration until the CDR. Each subsystem, in special the payloads, needs to be validated. In terms of interoperability, behaviour and requirements conformity, all the payloads engineering models must present a certain level of acceptance from the OBDH point of view. The HIL environment is a great opportunity to isolate the possible sources of failures, once each payload and even the OBDH can be tested on 1-by-1 operation. The Test Reports from the MIL environment (SATS) should guide this phase and point out the non-conformities and misunderstandings of the development team (OBDH and payloads). Due to some lack of expertise of the NanoSatC-BR2 development team, the HIL approach was not that much exercised and the project was rushed to be ready for complete integration. We will discuss the EM and FM integration at the Section~\ref{sec:results}.

 
	\label{fig:sastVerification}
 


\section{Learning Lessons and Reduce Complexity of Verification Work}
\label{sec:results}

This section discusses the effect of the V\&V processes adopted by INPE for CubeSat standard nanosatellites projects, highlighting the changes introduced by the use of Model-Based approaches. The process is supported by ECSS standards, but it is affected by COTS at very low procurement cost and a simple standardized architecture. The way a space organization deals with downsizing its traditional V\&V process to carry on V\&V of nanosatellites is not obvious.

It is observed in the SiS V\&V process presented in Section  \ref{sec:cubesat} that the process applied to nanosatellites is incremental due to both the characteristics of the requirements that are more volatile and the greater importance of interoperability in the integration of the nanosatellites subsystems.
Due to the volatility of the requirements and the higher acceptable risk in the nanosatellites project, the closing of the review documents is quite delayed, reaching cases in which the release is made a couple of phases later in the development process (see Figure \ref{tab:reviewDeliverables}, the release of the Requirements Baseline and Interface Requirement Document) or even not being formally produced (in Software Development Plan and Technical Specification). However, problems are identified when we start to deal with the spacecraft integration, EM and FM. 

Both integration phases will point to all kind of misunderstandings, non-conformities and errors, even mechanical, electrical or logical, or all of them, at the same time. A great part of the documents are released during this phases as result of gaps found during the integration. Thus, many requirements and constraints are added to the project to fit the solutions -- reworking hardware/software development is more difficult than reworking a document -- instead of the solutions to fit the requirements and constraints. This ends up with lack of information on the documents, lack of traceability and vulnerabilities associated with conflicting implementations.

Regarding the SiS discipline, the reduced complexity of verification of nanosatellites has been obtained by INPE with the lessons learned to support an automatic test generation adopted in  the V\&V process of traditional satellites. Although the  modelling activities are very time consuming, the benefits of the effective test suite provided allows INPE to increase confidence in model-based approaches. However, instead of effort in acceptance testing, which does not make sense in the nanosatellite projects, INPE focused on the use of modeling approaches to anticipate the detection and treatment of embedded software-intensive communicating subsystems in the development life cycle of SiS. The goal is to avoid software rework due to interoperability faults usually evidenced only in the integration testing.

The EM integration is aimed at the verification of every software subsystem  interoperability, beyond the mechanical and electrical integration. Being the OBDH the core of the satellite coordination, all the EM integration tests focus on how the OBDH software deals with other subsystems, centralizing their control and monitoring. The Test Cases used at MIL and HIL environments can be used at this point as a guide for the expected behaviour of the OBDH and its payloads. In this case, the models can be used as the single source of truth (SSOT). However, to ensure the closing of the EM integration phase, the space-to-ground integration must be addressed.

Not only the spacecraft must be working properly on the inside, it must be controllable from the ground. The model-based approach can be useful here, as the concept of spacecraft operation can be modelled in a higher level than the way it was done before with the subsystems. However, between the space and ground segments there is a communication channel that cannot be modelled in the same way preventing us to directly model the validation of interoperability between them. So, we use both the EM model and the Satellite Control Software directly to verify the complete operation of the spacecraft and exercise possible flight plans for the mission.

Ending the CDR, the FM integration can start. This phase does not differ much from the EM integration as we are trying to replicate the expected behaviour found at the EM. The main differences occur in the replacement of the subsystem engineering models by their flight models. Some of them are more fragile (e.g. solar panels instead of dummy Printed Circuit Boards - PCBs) or more robust (e.g. aluminium structure instead of a mock-up). The possible problems that can be found here are more related to the assembly and cabling, items that it is not possible to model or measure in the initial phases. At this stage, the chance of finding logical errors in the spacecraft software and its operation is lower.

For the NanoSatC-BR2 mission, all these lessons were pointed out. They allowed us to learn how to reduce the complexity of the V\&V process, in qualitative and quantitative ways. The gaps in the team expertise is difficult to by-pass as, most of time, the missions are university-class projects.

Therefore, INPE has made progress on research that contributes with a model-based approach in terms of (i)  alternative modeling formalisms \cite{almeida2017modeling}; (ii) new tools to aim deriving test cases from  models; (iii) open architectures  to support the test cases execution. A Scalable Architecture Test System (SATS) based on Arduino computer boards is proposed in \cite{approach} 
. It allows to systematize test scenarios based on the concept of model and hardware in the loop. The approach combines Model-Based Testing (MBT) and Model-Driven Engineering (MDE) techniques to support: (i) specification of interoperability requirements, whose behavior is modeled in state machines; (ii) automated software code generation from those models to be embedded in Arduino computer boards; (iii) automated generation of nominal test cases focusing on interoperability properties; (iv) extension of models with robustness aspects; and finally (v) automated generation of test cases focusing on robustness properties. Moreover, a failure emulator mechanism framework (FEM) is proposed in 
\cite{batista2019towards} for robustness testing of interoperable software-intensive subsystems on-board nanosatellite. FEM acts in the communication channel being part of the integration test workbench in two phases of nanosatellite design: (i) robustness requirement specification using model in the loop (MIL) and (ii) robustness validation using hardware in the loop (HIL). Both approaches have been applied in the V\&V process of NanosatC-BR2 \cite{almeida2017modeling}.

\section{Conclusions and future work}
\label{sec:conclusions}

This work presented the V\&V process, used at INPE, applied in nanosatellites projects that follow the CubeSat standard. The structure or even the subsystems of these nanosatellites are fully acquired as COTS. Therefore, the V\&V process usually adopted in traditional satellites, as recommended by ECSS standards, needs to be adapted to reflect the reduced complexity of work and the absence of one or more disciplines.

We presented the rationale to conceive a V\&V process for software-intensive subsystem on-board nanosatellites developed by INPE aiming to downsize the practices used in traditional satellite V\&V process. The rationale highlights the importance of the interoperability among the OBDH SiS and the satellite payloads. Being key elements, the success of the mission in orbit depends on confidence in their operation. 

Research opportunities were identified to improve the use of MBT approach, firstly applied in the process of the traditional satellites, and then to anticipate integration testing in the development life cycle of the Cubesat-based space missions using InRob. The development of standardized test environments using model-based methods and tools allowed the establishment of integration testing processes for satellite subsystems that are more effective in time and cost. The complexity of the subsystems,  increasing in functionalities implemented by software (SiS) is supported by the systematization of the generation and execution of interoperability and robustness tests provided by inRob, which has shown a strong alignment with the new philosophy of small satellite projects. So, standardisation, systematisation and automation of the V\&V process may help to work around problems (unknown  requirements  in  the  begin  of  the development cycle and their constant evolution, immaturity of the stakeholders and the lack of expertise of the development team regarding the specificities  of  software  engineering  for  space  area  and  low  budget  forcing  the  allocation  of  inexperienced developers  in  the  SiS  context) and reach a more compliant results. Furthermore, we believe that this V\&V process can have a broader impact to several embedded systems and infrastructures for critical machines.

In the near future, through scientific partnerships with national and international academic institutions, INPE hopes to consolidate the V\&V process, based on a test system that supports the approach of evolutionary integration tests for the development of SiS embedded in nanosatellites. The goal is to use the InRob method and associated tools at scale (model construction, automatic test generation and automated instrumentation in the execution of tests), systematizing the verification of the SiS shipped in the different nanosatellites that will constitute the new generation of The Brazilian Environmental Data Collecting Systems (BEDCS), in accordance with the CONASAT\footnote{http://www.crn.inpe.br/conasat1/nanosatt.php} project.

\section{Acknowledgment}

This work has been partially supported by the proj. ADVANCE - Addressing V\&V Challenges in Future Cyber-Physical Syst. (https://www.advance-rise.eu/ H2020-MSCA-RISE-2018, n. 823788) and partially supported by the Centro de Inform. e Sistemas da Univers. de Coimbra (CISUC).

\balance

\bibliographystyle{IEEEtran}
\bibliography{IEEEabrv,issre2020}

\end{document}